\documentclass[11pt,a4paper]{article}
\pdfoutput=1

\usepackage{jcappub}
\usepackage{tensor}

\usepackage[utf8]{inputenc}
\usepackage{graphicx}
\usepackage{verbatim}
\usepackage{epsfig}
\usepackage{subfigure}
\usepackage{color}
\usepackage{comment}
\usepackage{slashed}
\usepackage{amsmath}
\usepackage{ulem}
\usepackage{physics}

\newcommand{\be}{\begin{equation}} 
\newcommand{\ee}{\end{equation}}
\newcommand{\bea}{\begin{eqnarray}} 
\newcommand{\eea}{\end{eqnarray}}

\newcommand{\bmp}{\noindent\begin{minipage}{16cm}}
\newcommand{\emp}{\end{minipage}\vskip 7mm} 
\def\lsim{\mathrel{\raise.3ex\hbox{$<$\kern-.75em\lower1ex\hbox{$\sim$}}}}
\def\gsim{\mathrel{\raise.3ex\hbox{$>$\kern-.75em\lower1ex\hbox{$\sim$}}}}

\newcommand{\intron}[1]{}

\title{Initial conditions for plateau inflation: a case study}

\author[a]{Tommi Tenkanen}
\author[b,c]{and Eemeli Tomberg}

\affiliation[a]{\,Department of Physics and Astronomy, Johns Hopkins University, \\
3400 N. Charles Street, Baltimore, MD 21218, USA}
\affiliation[b]{\,Department of Physics, University of Helsinki and Helsinki Institute of Physics,\\
P.O. Box 64, FIN-00014 University of Helsinki, Finland}
\affiliation[c]{\,National Institute of Chemical Physics and Biophysics, \\
R{\"a}vala pst. 10, Tallinn, 10143, Estonia}

\emailAdd{ttenkan1@jhu.edu}
\emailAdd{eemeli.tomberg@helsinki.fi}

\abstract{We study initial conditions for inflation in scenarios where the inflaton potential has a plateau shape. Such models are those most favored by Planck data and can be obtained in a large number of model classes. As a representative example, we consider Higgs inflation with and without an $R^2$ term in the context of Palatini gravity. We show that inflation with a large number of e-folds generically occurs in a large part of the parameter space without any fine-tuning of parameters even when the scale of inflation and the inflaton field value during inflation are much smaller than the Planck scale. We discuss consequences for detection of primordial gravitational waves and spectral tilt of curvature perturbations, as well as the recently proposed "Trans-Planckian Censorship" conjecture.
}

\begin{document}

\begin{flushleft}
	\hfill		 HIP-2020-2/TH \\
\end{flushleft}

\maketitle



\section{Introduction}
\label{introduction}

In this paper we study the initial conditions that are needed to start a phase of accelerated expansion, i.e. cosmic inflation, in the case where inflation is driven by a scalar field whose potential exhibits a plateau regime. Scenarios like this are encountered, for example, in various models where the Standard Model (SM) Higgs field drives inflation \cite{Bezrukov:2007ep,Bauer:2008zj} and which are therefore of particular interest in the search for the connection between low energy particle physics and cosmic inflation (see also Refs. \cite{Spokoiny:1984bd,Futamase:1987ua,Salopek:1988qh,Fakir:1990eg,Amendola:1990nn,Kaiser:1994vs,CervantesCota:1995tz,Komatsu:1999mt} for earlier work on the topic and Refs. \cite{Rubio:2018ogq,Tenkanen:2020dge} for recent reviews), or in the context of so-called $\alpha$-attractor models \cite{Ferrara:2013rsa,Kallosh:2013yoa}. Furthermore, plateau models are precisely those most favored by the Cosmic Microwave Background (CMB) data \cite{Akrami:2018odb,Martin:2013nzq}.

In order for (slow-roll) inflation to happen, the kinetic energy stored in the inflaton field must remain (much) smaller than the potential energy. However, it is a non-trivial question of how generically this happens in a given model of inflation. This was studied in the seminal papers \cite{Linde:1985ub,Kung:1990,Goldwirth:1990,Goldwirth:1992} (see also Refs. \cite{Feldman:1989,East:2015ggf,Kleban:2016sqm,Clough:2016ymm,Linde:2017pwt,Azhar:2018nol,Bloomfield:2019rbs}), in which it was shown that "chaotic" models of inflation where inflation occurs at $\phi > M_{\rm P}$ are not very sensitive to the initial conditions but the occurence of inflation in the so-called small-field or hilltop models with $\phi \lesssim M_{\rm P}$ can depend heavily on the initial conditions. However, as recently shown in Ref. \cite{Chowdhury:2019otk}, models where the inflaton potential exhibits a plateau are not "small-field" models in the usual sense, and inflation can occur also at $\phi \sim M_{\rm P}$ without a need for fine-tuning.\footnote{Like Ref. \cite{Chowdhury:2019otk}, by "fine-tuned" initial conditions we mean a set of initial conditions that occupies only a small fraction of the inflaton phase-space. For discussion on the choice of a phase-space measure, see Ref. \cite{Remmen:2013eja} and Section VII of Ref. \cite{Chowdhury:2019otk}.}

In this study, we will further consolidate this claim by studying inflation with a non-trivial gravitational sector in the Palatini formulation of gravity, where the metric tensor and space-time connection are treated as independent variables. The models we consider have a plateau potential in the Einstein frame and a possibility for non-canonical kinetic terms but do not introduce new propagating fields, allowing us to study a wide range of possibilities in a simple setup. Moreover, the field range in these models can be considerably smaller than the Planck scale.

In this way, we extend the studies in Refs. \cite{Guth:2013sya,Chowdhury:2019otk} to address the claims made by Ijjas et al. in Ref. \cite{Ijjas:2013vea} concerning fine-tuning in the initial conditions needed to start inflation and how the Planck data favor models for which this issue is, according to Ref. \cite{Ijjas:2013vea}, most problematic. We will show that this is generically not the case in a homogeneous and isotropic situation even in scenarios where inflation occurs at very small field values, $\phi \ll M_{\rm P}$. In addition to presenting this important result, we will also comment on the claims made in Ref. \cite{Bedroya:2019tba} regarding the required amount of fine-tuning in initial conditions for inflation in the context of the "Trans-Planckian Censorship" conjecture and show that generically no fine-tuning is needed in models like those studied in this paper. 

The paper is organized as follows: in Sec. \ref{inflation}, we present the class of inflationary models we study in this paper and discuss the scaling laws that the inflaton equation of motion and the Friedmann equation exhibit in our scenario. In Sec. \ref{results}, we present our main results regarding the initial conditions and predictions for CMB observables in these models and then conclude in Sec. \ref{conclusions}.


\section{Plateau inflation}
\label{inflation}

In models of plateau inflation, the inflaton potential approaches a constant for large field values. The exact form of the potential varies from model to model, but the basic features remain the same. For high energy densities, kinetic energy of the field always dominates over the potential term, since the potential is bounded from above. However, when kinetic energy is low enough, the potential dominates, and there is a possibility for slow-roll inflation to happen. The inflationary stage ends in (p)reheating once the field rolls close to the potential minimum.  We consider the model presented below to be a representative example of this behaviour.

We start with the action
\begin{equation}
\label{SJ}
S_J = \int {\rm d}^4x \sqrt{-g}\left(\frac12 F(R,\phi)- \frac12g^{\mu\nu}\partial_\mu\phi\partial_\nu\phi - V(\phi)\right) ,
\end{equation}
where $g$ is the determinant of the space-time metric $g_{\mu\nu}$, $R=g^{\mu\nu}R_{\mu\nu}(\Gamma,\partial\Gamma)$ is the curvature scalar, $R_{\mu\nu}$ is the Ricci tensor which is constructed by contraction from the Riemann tensor $R\indices{^\lambda_{\mu\nu\sigma}}$ which, in turn, is constructed from the space-time connection $\Gamma$ and its first derivatives, and $\phi$ is the inflaton field. We take
\begin{equation}
\label{F}
F(R,\phi) = M_{\rm P}^2 R + \alpha R^2 +G(\phi)R,
\end{equation}
where $\alpha$ is a dimensionless parameter and $G(\phi)$ encapsulates the possible non-minimal couplings between the inflaton and gravity.

We use the Palatini or metric-affine formulation of general relativity, where the metric $g_{\mu\nu}$ and the connection $\Gamma$ are taken to be independent variables, with the requirement that the connection is symmetric in its two lower indices, $\Gamma^\lambda_{\mu\nu}=\Gamma^\lambda_{\nu\mu}$, so that the torsion tensor vanishes (for scenarios with non-vanishing torsion, see Refs. \cite{Rasanen:2018ihz,Shimada:2018lnm}). When $F(R,\phi)$ in \eqref{F} is not of the standard Einstein-Hilbert form, the connection will not be of the usual Levi-Civita built solely from $g_{\mu\nu}$ but also depends on $\phi$. Note that this is only important during inflation: in the late universe, the extra terms in \eqref{F} are insignificantly small for all parameter values considered in this paper, and the Palatini formulation will give predictions which are indistinguishable from the usual metric formulation. However, the benefits of this formulation for our purpose will become evident below. For references on the Palatini formulation, see Ref. \cite{Bauer:2008zj} for original work and Ref. \cite{Tenkanen:2020dge} for an introduction to the topic, as well as Refs. \cite{Bauer:2010jg,Tamanini:2010uq,Enqvist:2011qm,Azri:2017uor,Rasanen:2017ivk,Tenkanen:2017jih,Racioppi:2017spw,Markkanen:2017tun,Jarv:2017azx,Racioppi:2018zoy,Enckell:2018kkc,Carrilho:2018ffi,Aoki:2018lwx,Enckell:2018hmo,Antoniadis:2018ywb,Rasanen:2018fom,Kannike:2018zwn,Rasanen:2018ihz,Almeida:2018oid,Antoniadis:2018yfq,Shimada:2018lnm,Takahashi:2018brt,Jinno:2018jei,Tenkanen:2019jiq,Rubio:2019ypq,Jinno:2019und,Tenkanen:2019xzn,Tenkanen:2019wsd,Gialamas:2019nly,Racioppi:2019jsp,Shaposhnikov:2020geh} for some other recent studies on inflation in this context. In the following, we will take $G(\phi)=\xi \phi^2$ for simplicity. Then $\alpha=0\, , \, \xi\neq 0$ corresponds to the usual non-minimal inflation (see e.g. Refs. \cite{Spokoiny:1984bd,Futamase:1987ua,Salopek:1988qh,Fakir:1990eg,Amendola:1990nn,Kaiser:1994vs,Komatsu:1999mt,Bezrukov:2007ep,Barvinsky:2008ia,Lerner:2009xg,Kaiser:2013sna,Kallosh:2013maa,Kallosh:2013tua,Chiba:2014sva,Boubekeur:2015xza}) and $\alpha\neq 0$ to the scenario studied in Refs. \cite{Enckell:2018hmo,Antoniadis:2018ywb,Antoniadis:2018yfq,Tenkanen:2019jiq,Tenkanen:2019wsd,Gialamas:2019nly} (for studies in the metric case, see Refs. \cite{Salvio:2015kka,Calmet:2016fsr,Wang:2017fuy,Ema:2017rqn,He:2018gyf,Ghilencea:2018rqg,Gundhi:2018wyz,Karam:2018mft,Enckell:2018uic,Canko:2019mud}), while $\alpha=0=\xi$ represents the minimally coupled case where only the form of $V(\phi)$ is important.

We use standard methods to proceed from \eqref{SJ}: we eliminate the $R^2$ term in favour of an auxiliary field $\varphi\equiv M_{\rm P}^2 + 2\alpha R$, perform a Weyl transformation 
\begin{equation}
g_{\mu\nu}\to \frac{\varphi + G(\phi)}{M_{\rm P}^2} g_{\mu\nu} ,
\end{equation}
eliminate the auxiliary field $\varphi$, and re-define the inflaton field, ending up with the Einstein frame action \cite{Enckell:2018hmo}
\begin{equation}
\label{SE}
S_E = \int {\rm d}^4x \sqrt{-g}\bigg[\frac12 M_{\rm P}^2 R- \frac12\partial^\mu\chi\partial_\mu\chi
+ \frac{\alpha}{2 M_{\rm P}^4}\left(1+8\alpha\frac{\bar{U}}{M_{\rm P}^4}\right)\left(\partial^\mu\chi\partial_\mu\chi\right)^2 - U(\chi)\bigg] .
\end{equation}
The re-defined inflaton field $\chi$ is given by
\begin{equation} \label{field_redefinition}
\frac{{\rm d}\phi}{{\rm d}\chi} =\sqrt{\left(1+\frac{G(\phi)}{M_{\rm P}^2}\right)\left(1+8\alpha\frac{\bar{U}}{M_{\rm P}^4}\right)} ,
\end{equation}
and the potentials are
\begin{equation} \label{U}
U(\chi) \equiv \frac{\bar{U}(\chi)}{1+8\alpha\bar{U}(\chi)/M_{\rm P}^4}\,, \quad
\bar{U}(\chi) \equiv \frac{V(\phi(\chi))}{\qty[1+G(\phi(\chi))/M_{\rm P}^2 ]^2} .
\end{equation}
When $\bar{U}$ either grows or approaches a constant, the potential $U$ indeed exhibits a plateau.

Note that in the Palatini formulation, only one field is dynamical; this is related to the fact that the Jordan frame connection depends on both $\phi$ and $g_{\mu\nu}$ and not just the metric. This renders the $\alpha R^2$ term non-dynamical, unlike in the case of the famous Starobinsky model~\cite{Starobinsky:1980te}, which is based on metric gravity (where the connection is precisely the Levi-Civita one). However, a non-zero $\alpha R^2$ term will still affect the dynamics of $\chi$, as we will see more explicitly in the following. We then have a plateau model with only one dynamical field but significant freedom to choose $\alpha$, $G(\phi)$, and $V(\phi)$, which allows us to e.g. build models with very small inflationary field values, as we will see in Sec. \ref{cmb_predictions}.

In the Friedmann–Robertson–Walker (FRW) case with zero spatial curvature the Friedmann equation and the equation of motion for the field read 
\cite{Enckell:2018hmo}
\begin{eqnarray} \label{FRW_eqs}
3 M_{\rm P}^2 H^2 &=& \frac12 \bigg[1+3\alpha(1+8\alpha\frac{\bar{U}}{M_{\rm P}^4})\frac{\dot{\chi}^2}{M_{\rm P}^4} \bigg]\dot{\chi}^2+U\,,\\ \nonumber
0 &=& \bigg[1+6\alpha(1+8\alpha\frac{\bar{U}}{M_{\rm P}^4})\frac{\dot{\chi}^2}{M_{\rm P}^4}\bigg]\ddot{\chi}+3\bigg[1+2\alpha(1+8\alpha\frac{\bar{U}}{M_{\rm P}^4})\frac{\dot{\chi}^2}{M_{\rm P}^4}\bigg]H\dot{\chi} + 12\alpha^2\frac{\dot{\chi}^4}{M_{\rm P}^8}\bar{U}' + U'\, ,
\end{eqnarray}
and inflation happens when the first slow-roll parameter
\begin{equation} \label{epsilon_H}
	\epsilon_H \equiv -\frac{\dot{H}}{H^2} = \frac{\dot{\chi}^2}{2M_{\rm P}^2H^2}\qty[ 1 + 2\alpha\qty(1+8\alpha\frac{\bar{U}}{M_{\rm P}^4})\frac{\dot{\chi}^2}{M_{\rm P}^4} ]
\end{equation}
is smaller than one.

\begin{figure}
\centering
\includegraphics[scale=1]{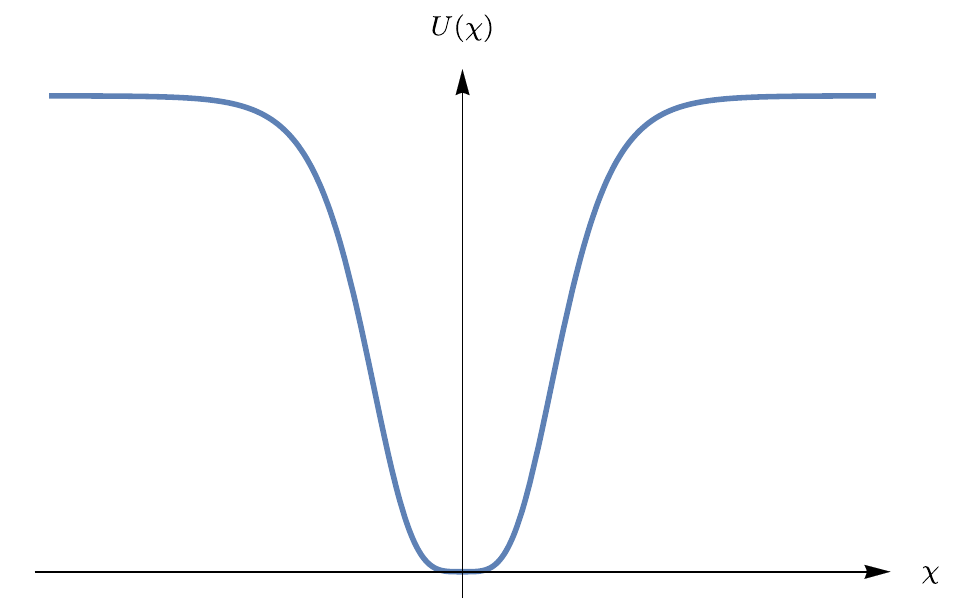}
\caption{A sketch of the Einstein frame potential $U(\chi)$ \eqref{U} with the quartic Jordan frame potential \eqref{V}. The potential is exponentially flat for $\chi/M_{\rm P} > (\xi^2 + 2\alpha_{\rm eff})^{-1/4}$.
}
\label{plateau_potential}
\end{figure}
In Ref. \cite{Enckell:2018hmo}, the equations \eqref{FRW_eqs} were studied in the slow-roll limit, where $\epsilon_H \ll 1$ and the $\ddot{\chi}$-term in \eqref{FRW_eqs} is negligible. It was shown that the leading slow-roll predictions for the power spectrum amplitude and the spectral index do not depend on $\alpha$ but are equal to the standard $\alpha=0$ case, that is,
\begin{equation} \label{As_and_ns}
	24 \pi^2 M_{\rm P}^4 A_s = \frac{U}{\epsilon_U} = \frac{\bar{U}}{\epsilon_{\bar{U}}} \, , \qquad n_s = 1 + 2\eta_U - 6\epsilon_U = 1 + 2\eta_{\bar{U}} - 6\epsilon_{\bar{U}} \, ,
\end{equation}
where the potential slow-roll parameters are defined as
\begin{equation}
\begin{aligned}
	\epsilon_U &\equiv \frac{1}{2}M_{\rm P}^2\qty(\frac{U'}{U})^2 \, , \qquad
	& \epsilon_{\bar{U}} &\equiv \frac{1}{2}M_{\rm P}^2\qty(\frac{\bar{U}'}{\bar{U}})^2 = \epsilon_U \large|_{\alpha=0} \, , \\
	\eta_U &\equiv M_{\rm P}^2\frac{U''}{U} \, , \qquad
	& \eta_{\bar{U}} &\equiv M_{\rm P}^2\frac{\bar{U}''}{\bar{U}} = \eta_U\large|_{\alpha=0} \, .
\end{aligned}
\end{equation}
The number of e-folds of inflation does not depend on $\alpha$ either.\footnote{When the quantities with and without a bar are compared here, it is done with a fixed value of the Jordan frame field $\phi$.} The main effect of a non-zero $\alpha$ is to lower the predicted tensor-to-scalar ratio:
\begin{equation} \label{r}
	r = 16\epsilon = \frac{16}{1+8\alpha\bar{U}/M_{\rm P}^4}\epsilon_{\bar{U}} \, ,
\end{equation}
so that in case of large $\alpha$, the scenario yields negligible primordial gravitation waves. The Planck and BICEP2/Keck Array limits for the aforementioned observables are \cite{Akrami:2018odb,Ade:2018gkx}
\begin{equation} \label{planck}
	A_s = 2.1 \times 10^{-9} \, , \quad n_s = 0.9625 \pm 0.0048 \, , \quad r < 0.06 \, ,
\end{equation}
measured at the pivot scale $k=0.05\,{\rm Mpc}^{-1}$. In this paper, we will show by studying a few example potentials that the slow-roll solution is still an attractor in the presence of the non-minimal kinetic terms, so the results \eqref{As_and_ns} and \eqref{r} are general and practically independent of the initial conditions.

\subsection{An example: Higgs-like inflation}
As a representative example, we study the potential
\begin{equation} \label{V}
	V(\phi) = \frac{\lambda}{4}\phi^4 \, ,
\end{equation}
which is the (Jordan frame) potential encountered in e.g. the Higgs inflation model, where the SM Higgs is the field responsible for driving inflation \cite{Bezrukov:2007ep,Bauer:2008zj}. For studies on initial conditions in the metric case, see Refs. \cite{Salvio:2015kka,Salvio:2017oyf,Mishra:2018dtg,Mishra:2019ymr}. In the Palatini case, the potential $U(\chi)$ is exponentially flat for large field values as long as either $\xi$ or $\alpha$ is non-zero, see Fig. \ref{plateau_potential}. Explicitly, in terms of $\phi$,
\begin{equation} \label{U_Higgs}
	U(\phi) = \frac{\lambda \phi^4}{4\qty[\phi^4\qty(\xi^2 + 2\alpha\lambda)/M_{\rm P}^4 +  2\xi\phi^2/M_{\rm P}^2 + 1 ]} \, ,
\end{equation}
which approaches a constant for $\phi/M_{\rm P} > (\xi^2 + 2\alpha\lambda)^{-1/4}$, with $\chi \approx \phi$ still at the starting point of the plateau.

If $\xi \ll 1$, we have slow-roll inflation with the quartic potential \eqref{V} at the CMB scales, and a non-zero $\alpha$ only modifies $r$ as discussed above. The observables in terms of the number of e-folds $N$ are easily calculated and read
\begin{equation} \label{phi4_observables}
	A_s = \frac{2\lambda N^3}{3\pi^2} \, , \quad n_s = 1 - \frac{3}{N} \, , \quad r = \frac{16}{N(1 + 128\alpha\lambda N^2)} \, .
\end{equation}
If $\xi \gg 1$, we have the Palatini-Higgs inflation \cite{Bauer:2008zj}, so
\begin{equation} \label{higgs_inf_observables}
	A_s = \frac{\lambda N^2}{12\pi^2 \xi} \, , \quad n_s = 1 - \frac{2}{N} \, , \quad r = \frac{2}{\xi N^2(1 + 2\alpha\lambda/\xi^2)} \, .
\end{equation}
Here $N$ depends on the energy scale of inflation, and using the observed value of $A_s$ and assuming instant reheating it can be written for our pivot scale $k=0.05\,{\rm Mpc}^{-1}$ as
\begin{equation} \label{N}
	N \approx 56 - \frac{1}{4}{\rm ln}\left( \frac{0.06}{r}\right) \, .
\end{equation}
We will consider the compatibility of these predictions with the Planck observations \eqref{planck} in section \ref{cmb_predictions}.

Before analyzing the time evolution in this model, we note that the model obeys a scaling law: when the coordinates and couplings in the Jordan frame action \eqref{SJ} are scaled as
\begin{equation} \label{scaling}
	x^\mu \to sx^\mu \, , \quad \alpha \to s^2\alpha \, , \quad \lambda \to s^{-2}\lambda \, ,
\end{equation}
where $s$ is a scaling parameter, then the scalar curvature changes as $R \to s^{-2} R$, and the action stays the same apart from a constant scaling:\footnote{Note that this does not depend on the form of $V(\phi)$; in general, the $\lambda$-scaling can be replaced by a scaling of $V(\phi)$.}
\begin{equation} \label{scaling_SJ}
	S_J \to s^2 S_J \, .
\end{equation}
In this scaling, the parameter
\begin{equation} \label{alpha_eff}
	\alpha_{\rm eff} \equiv \alpha\lambda
\end{equation}
remains constant. The classical equations of motion do not change in such a rescaling of the action. There is then no change in some of the dimensionless parameters, such as the slow-roll parameters, $r$, or $n_s$ as functions of the number of e-folds of inflation, as can be directly seen from \eqref{phi4_observables}, \eqref{higgs_inf_observables}. There is a scaling in the dimensionful parameters, however, such as the energy scale of inflation, as well as in $A_s$. 

In the next section, we will consider the time evolution of the model in phase-space with different parameter values. Because of the scaling symmetry, it is enough to consider different values of $\xi$ and $\alpha_{\rm eff}$. The phase-space flow diagrams we present do not depend on the values of $\lambda$ and $\alpha$ separately. These can always be fixed for any $\alpha_{\rm eff}$ by, for example, fixing the power spectrum amplitude $A_s$ at the CMB scale.


\section{Results}
\label{results}

\subsection{Attractor behaviour}
\label{attractor}

\begin{figure}
\centering
\includegraphics[scale=1]{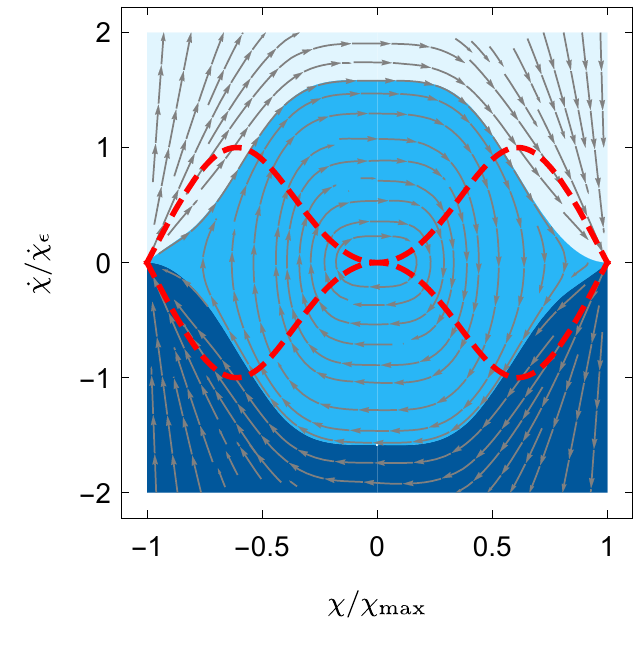}
\hspace{1cm}
\includegraphics[scale=1]{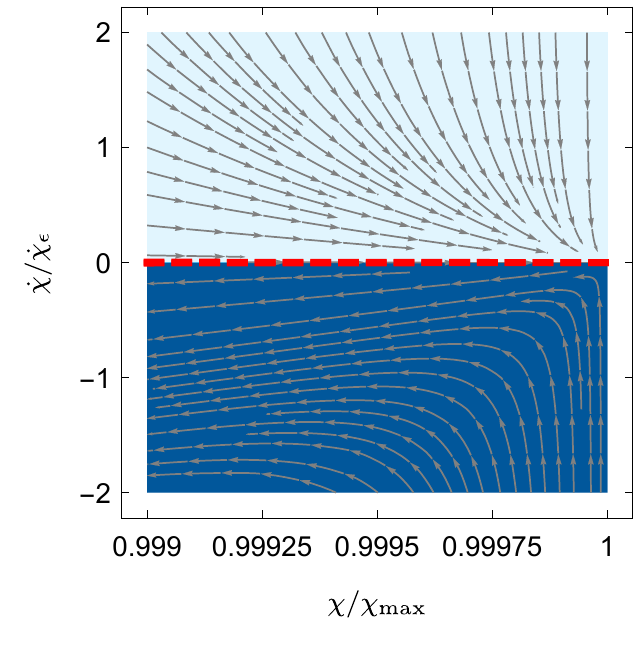}
\caption{Phase-space flow of the solutions to Eq. \eqref{FRW_eqs} with $\xi=0$, $\alpha_{\rm eff}=10^9$. The red dashed lines correspond to $\epsilon_H = 1$, so inflation occurs in points between them. Slow-roll inflation happens near the points $(\pm\chi_{\rm max},0)$. Trajectories in the light-coloured upper region eventually end up to the slow-roll regime at $\chi \approx \chi_{\rm max}$ and the trajectories in the dark-coloured lower region eventually end up to the slow-roll regime at $\chi \approx -\chi_{\rm max}$. From there, after a long slow-roll period, the field enters the middle region and starts to oscillate around the potential minimum. Only trajectories which start in this middle region never enter slow-roll. \textbf{Left:} The whole $\chi$-range, with $\dot{\chi}_\epsilon$ defined to be the maximum $\dot{\chi}$-value on the dashed lines. \textbf{Right:} A zoomed-in version near the positive $\chi_{\rm max}$ showing the convergence of trajectories to the horizontal slow-roll line from above, and the divergence of trajectories below which will overshoot the minimum and end up into slow-roll on the other side of the potential.}
\label{alpha_10to9_chi}
\end{figure}

\begin{figure}
\begin{center}
\includegraphics[scale=0.75]{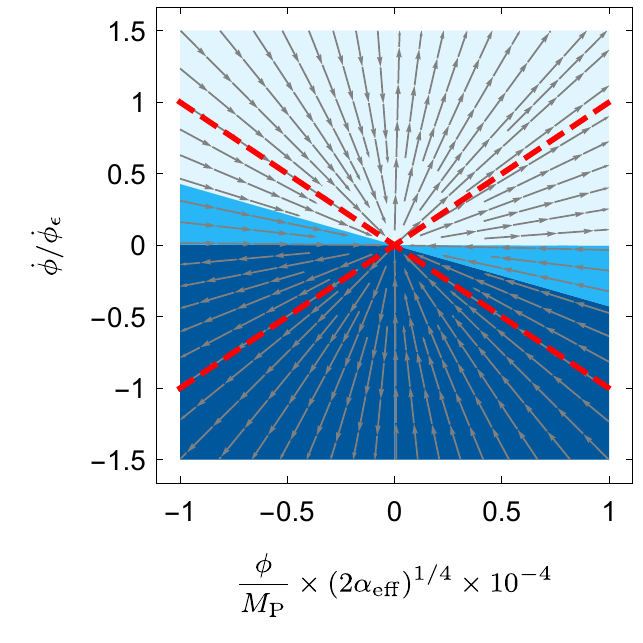}
\includegraphics[scale=0.75]{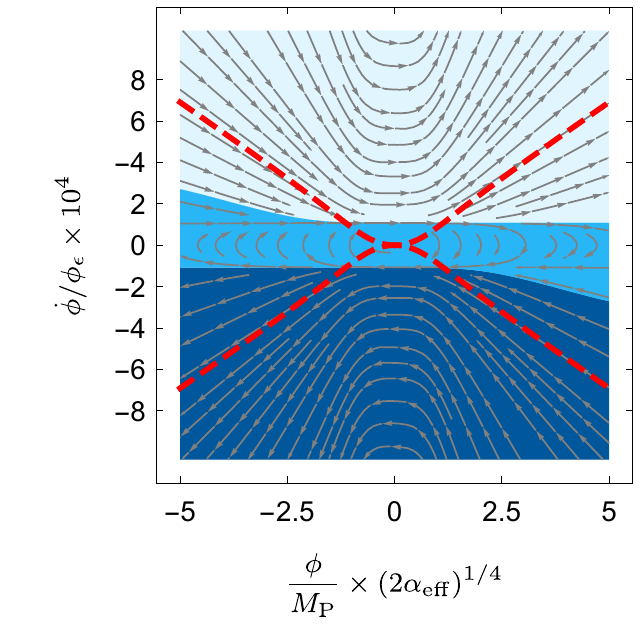}
\includegraphics[scale=0.75]{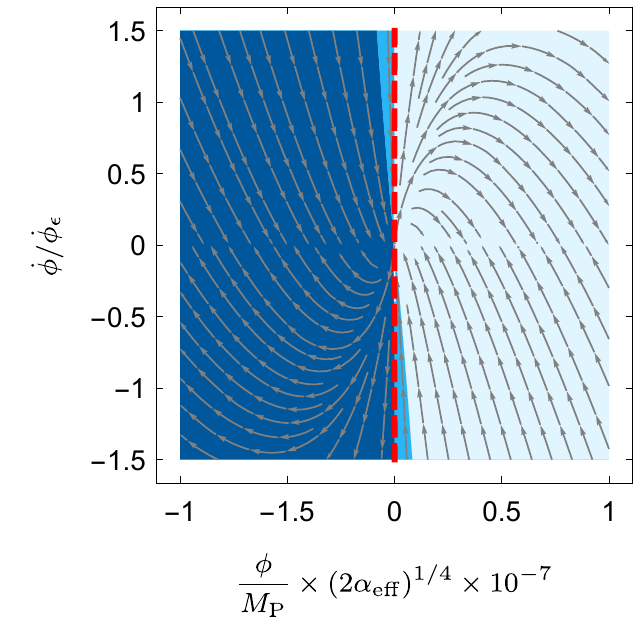}
\end{center}
\vspace{-0.5cm}
\caption{Same as Fig. \ref{alpha_10to9_chi} but with the Jordan frame field $\phi$ instead of $\chi$. Slow-roll happens near $\dot{\phi}=0$, and $\dot{\phi}_\epsilon$ is defined as the $\dot{\phi}$-value with $\epsilon_H=1$ when $\phi=10^4 M_{\rm P} \times (2\alpha_{\rm eff})^{-1/4}$. \textbf{Left:} A representative image of the phase-space, with the maximum value of $\phi$ chosen so that it corresponds to $\mathcal{O}(100)$ e-folds of slow-roll inflation. \textbf{Middle:} A zoomed-in version showing the flow near the origin. \textbf{Right:} A zoomed-out version, which shows the eventual convergence of trajectories towards slow-roll.}
\label{alpha_10to9_phi}
\end{figure}

We study the phase-space flow of solutions to \eqref{FRW_eqs} with the potential \eqref{V} in two cases. First, Figs. \ref{alpha_10to9_chi} and \ref{alpha_10to9_phi} show the inflaton trajectories in phase-space for the parameter values 
\begin{equation}
	\xi=0 \, , \qquad \alpha_{\rm eff}=10^9 \, .
\end{equation}
As emphasized above, these trajectories only depend on $\alpha_{\rm eff}$, not $\alpha$ and $\lambda$ separately. For this choice, the canonical field $\chi$ has a maximum value
\begin{equation}
	\chi_{\rm max} \approx 8.9 \times 10^{-3} M_{\rm P} \,,
\end{equation}
corresponding to $\phi \to \infty$, which comes from integrating \eqref{field_redefinition} with $G=0$. The inflationary plateau is compressed to values close to $\chi_{\rm max}$, so the phase-space diagrams in Fig. \ref{alpha_10to9_chi} presented in $(\chi,\dot{\chi})$ coordinates are not very informative. For this reason, Fig. \ref{alpha_10to9_phi} shows the diagrams in $(\phi, \dot{\phi})$ coordinates with a long, explicit plateau at $\phi \gg M_{\rm P}(2\alpha_{\rm eff})^{-1/4}$.

\begin{figure}
\centering
\includegraphics[scale=1]{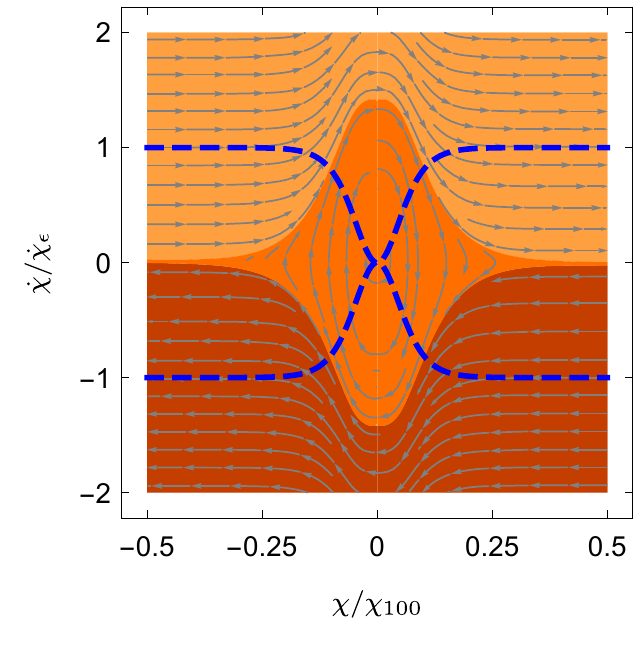}
\hspace{1cm}
\includegraphics[scale=1]{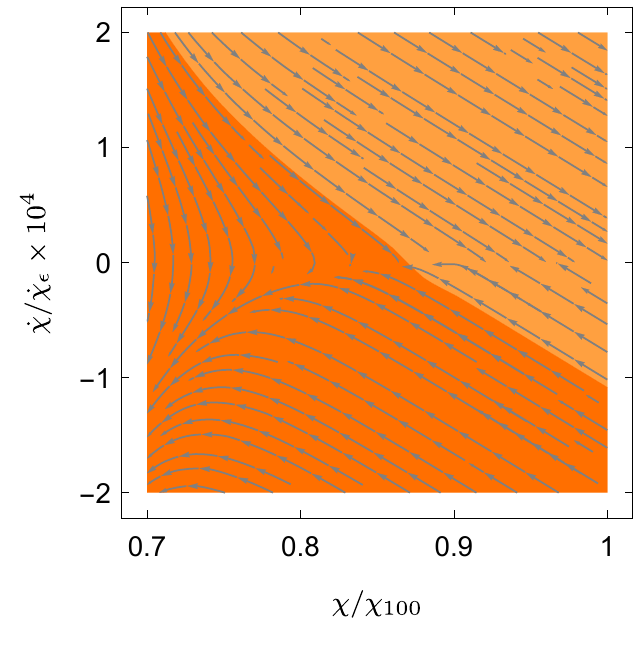}
\caption{Same as Fig. \ref{alpha_10to9_chi} but for $\alpha_{\rm eff} = 0$, $\xi = 9.5 \times 10^8$. Here $\chi_{100}$ is the minimum value of $\chi$ where there is $\mathcal{O}(100)$ e-folds of slow-roll inflation left. \textbf{Left:} A wide view; the trajectories continue almost horizontally outside of the figure. \textbf{Right:} A zoomed-in version near $\chi_{100}$, where the curving of the trajectories to the slow-roll attractor near $\dot{\chi}=0$ can be seen.}
\label{xi_10to9_chi}
\end{figure}

\begin{figure}
\begin{center}
\includegraphics[scale=0.75]{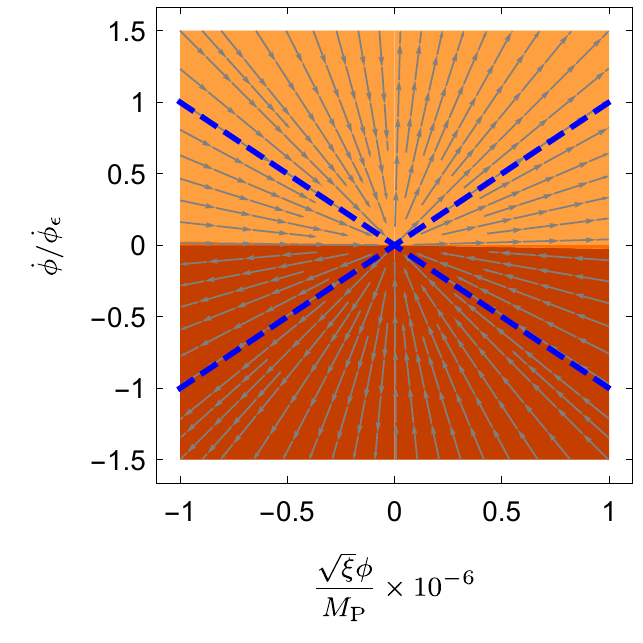}
\includegraphics[scale=0.75]{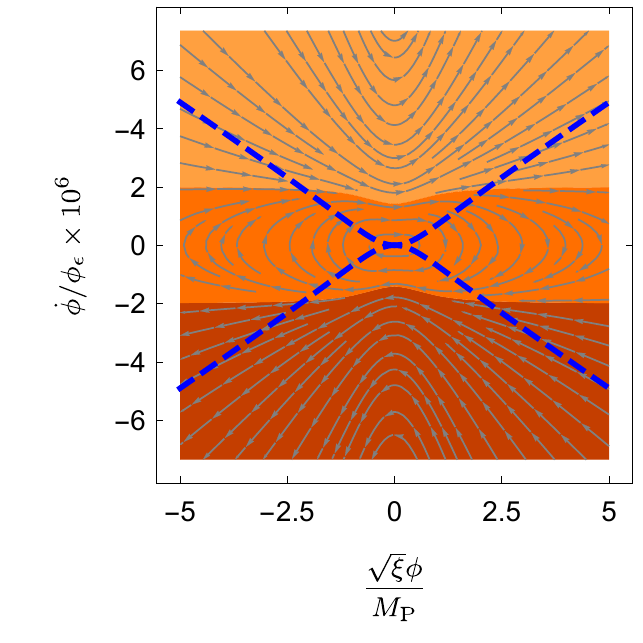}
\includegraphics[scale=0.75]{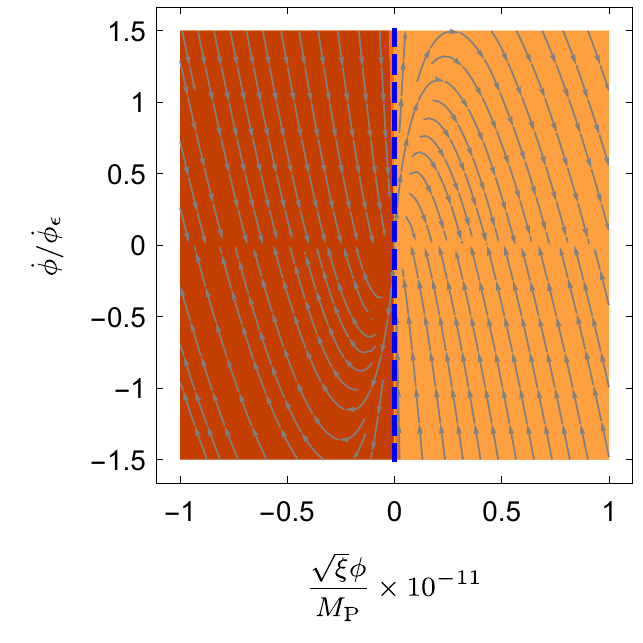}
\end{center}
\vspace{-0.5cm}
\caption{Same as Fig. \ref{alpha_10to9_phi} but for $\alpha_{\rm eff} = 0$, $\xi = 9.5 \times 10^8$. The field value $\phi = 10^6 M_{\rm P}/\sqrt{\xi}$ corresponds to $\mathcal{O}(100)$ e-folds of slow-roll inflation and $\dot{\phi}_\epsilon$ is the $\dot{\phi}$-value with $\epsilon_H=1$ when $\phi = 10^6 M_{\rm P}/\sqrt{\xi}$. The behaviour is almost identical to Fig. \ref{alpha_10to9_phi} despite the fact that the middle region without slow-roll is almost too narrow to be seen in the left and right panels.}
\label{xi_10to9_phi}
\end{figure}

Second, Figs. \ref{xi_10to9_chi} and \ref{xi_10to9_phi} show the trajectories for
\begin{equation}
	\xi=9.5\times 10^8 \, , \qquad \alpha_{\rm eff}=0 \, ,
\end{equation}
which gives the correct CMB predictions for $\lambda=0.1$, see Eqs. \eqref{planck}, \eqref{higgs_inf_observables}. This time, there is no maximum value for $\chi$ but at the CMB scales, $N \approx 50$, we have $\phi \gg M_{\rm P}$ and the canonical field value
\begin{equation}
	\chi_{50} \approx 4.6 \times 10^{-4} M_{\rm P} \, ,
\end{equation}
well below the Planck scale.

As can be seen from the figures, for most trajectories, slow-roll inflation will eventually take place. These trajectories either fall into slow-roll directly, or if they start with enough kinetic energy, they roll up the potential (possibly crossing $\chi=0$ first), slow down, and enter slow-roll when the field velocity has turned around. In the end, all trajectories roll down into the center region and start to oscillate around the potential minimum, transferring their energy density into other degrees of freedom through reheating, which is expected to be very efficient in our model, see Ref. \cite{Rubio:2019ypq}. Only trajectories with fine-tuned initial conditions end up in the oscillatory phase without going through a significant amount of slow-roll inflation first.

\pagebreak
Although we have depicted only two examples here, this behaviour is generically true for all potentials where at least one of the parameters $\xi$, $\alpha_{\rm eff}$ is large, assuming that every point in the $(\phi,\dot{\phi})$ or $(\chi,\dot{\chi})$ plane is equally likely as an initial condition. It is, of course, not clear if either of these probability measures is the correct one to use; however, it seems unlikely that the result would change by choosing any reasonable alternate probability measure. Thus, at least in a qualitative sense, we see that for most initial conditions, a long slow-roll period is expected.

\subsection{Predictions for observables}
\label{cmb_predictions}

Finally, let us see when the predictions of the model comply with the CMB observations \eqref{planck}, in particular with the measured value of $n_s$, as for large $\xi$ or $\alpha$ the tensor-to-scalar ratio \eqref{higgs_inf_observables} is always within the Planck bounds and the value of $\lambda$ can always be chosen such that the predicted amplitude $A_s$ matches with observations. From \eqref{N}, we see that for $N \lesssim 55$, the quartic case with small $\xi$ gives too small predictions for the spectral index regardless of the value of $\alpha_{\rm eff}$, $n_s \lesssim 0.945$ (see Eq. \eqref{phi4_observables}), which are ruled out by CMB observations. However, the large-$\xi$ case gives $n_s \lesssim 0.964$ (see Eq. \eqref{higgs_inf_observables}), which is compatible with observations. Note that the non-minimal coupling does not need to be very large, as we only need to require $\xi \gtrsim 1$. Therefore, for large enough $\xi$ and $N$, the Higgs-like $\phi^4$ model is generally compatible with observations.

\begin{figure}
\centering
\includegraphics[scale=1.5]{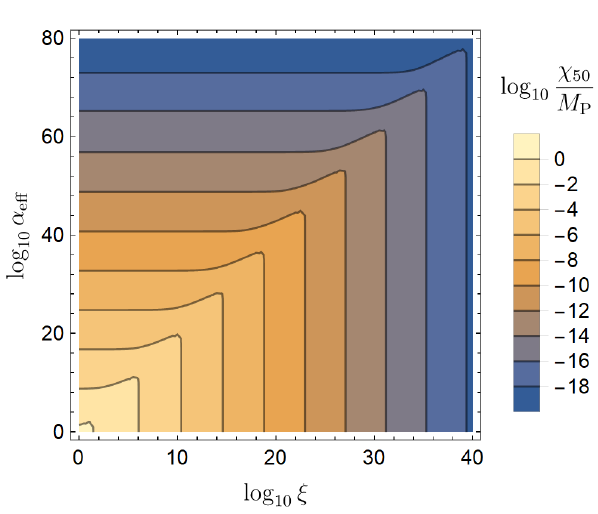}
\caption{The canonical field value $\chi$ at CMB scales corresponding to a time when there was 50 e-folds of inflation left, $\chi_{50}$, as a function of $\xi$ and $\alpha_{\rm eff}$. Roughly, the value of $\chi_{50}$ is dictated by $\xi^2 + 2\alpha_{\rm eff}$. Note that the field excursion during inflation, $\Delta \chi$, can be much smaller than the actual field value $\chi_{50}$, especially for $\alpha_{\rm eff} \gg \xi^2$, where all inflation happens close to a constant $\chi$-value, as discussed in section \ref{attractor}.
}
\label{alpha_xi}
\end{figure}

Let us then consider a case where the scale of inflation is small and where, consequently, the number of e-folds between the horizon exit of the pivot scale and the end of inflation is also small. This happens in scenarios with a large $\xi$ or $\alpha_{\rm eff}$, as then inflation takes place at small field values $\chi$ and the tensor-to-scalar ratio $r$ is small. To see how the value of $\chi$ at the CMB scales maps to $\xi$ and $\alpha_{\rm eff}$ in our model, see Fig. \ref{alpha_xi}. By choosing a large enough $\xi$ or $\alpha_{\rm eff}$, the excursion of the field during inflation $\Delta \chi$ and $r$ can be made small enough to agree with the so-called trans-Planckian censorship conjecture limits \cite{Bedroya:2019tba}
\begin{equation} \label{TCC_limits}
	r \lesssim 10^{-30},  \, \qquad |\Delta \chi| \lesssim 10^{-13}M_{\rm P}\,,
\end{equation}
needed to ensure that no scale observed in today's (classical) universe originated from super-Planckian energy scales during inflation. With Eqs. \eqref{phi4_observables}, \eqref{higgs_inf_observables} and $N \sim$ 50, these translate to
\begin{equation} \label{TCC_parameter_limits}
\begin{split}
	\alpha_{\rm eff} &\gtrsim 10^{24} \, , \quad \xi \ll 1 \, , \\
	\alpha_{\rm eff}/\xi &\gtrsim 10^{26} , \quad 1 \lesssim \xi^2 \ll \alpha_{\rm eff} \, , \\
	\xi &\gtrsim 10^{27} \, , \quad \alpha_{\rm eff} \ll \xi^2 .
\end{split}
\end{equation}
Demanding perturbativity, $\lambda < 1$, and the observed value for $A_s$ (see Eqs. \eqref{planck} and \eqref{higgs_inf_observables}), gives an upper limit for $\xi$\footnote{For Higgs inflation with the SM Higgs, more strict bounds may be necessary to match the electroweak scale measurements of $\lambda$; however, here we consider a general model with no bounds on $\lambda$ other than perturbativity.}:
\begin{equation} \label{xi_upper_limit}
	\xi \lesssim 10^{10} \, .
\end{equation}
Thus, to satisfy the limits \eqref{TCC_parameter_limits}, $\alpha_{\rm eff}$ must be large.

Note that when $r$ is very small, this starts to affect the CMB value of $N$, see Eq. \eqref{N}, and thus also the predicted value of $n_s$. In the limit $r<10^{-30}$, we have $N < 39$ and therefore $n_s < 0.95$ even in the case of a large $\xi$, in tension with observations. Therefore, while there is no problem with initial conditions, the Higgs-like inflation model with an $R^2$ term in Palatini gravity is in tension with observations if we also demand it to satisfy the limits \eqref{TCC_parameter_limits}.

However, for a shallower potential, for instance for $\phi^{2n}$ with $n<2$ and a non-minimal coupling $\xi\phi^n$, the prediction for the spectral index becomes $n_s = 1-(1+n/2)/N$ at the limit of large $\xi$ \cite{Jarv:2017azx}, which can be made compatible with Planck even when $N \lesssim 40$; see an example of this in Ref. \cite{Tenkanen:2019wsd}. Therefore, while the Higgs-like $\phi^4$ scenario is in tension with the CMB observations when requiring it to satisfy the proposed trans-Planckian censorship conjecture limits, this is not, in general, the case for scenarios belonging to the same model class. While we have not studied initial conditions for general $\phi^{2n}$ potentials in this paper, we do not expect them to differ from the Higgs-like $\phi^4$ case in scenarios where also an $R^2$ term in Palatini gravity is added to the Lagrangian.

Note also that, strictly speaking, the trans-Planckian censorship conjecture limits the length of the full inflationary period, not just inflation near the CMB scales. Even if the CMB scales satisfy the limits \eqref{TCC_limits}, the conjecture forbids too long prior periods of inflation. This may forbid a large section of the phase space in diagrams such as Figs. \ref{alpha_10to9_chi}, \ref{xi_10to9_chi}, \ref{alpha_10to9_phi}, and \ref{xi_10to9_phi}, though for big $\xi$ or $\alpha_{\rm eff}$, a large allowed region remains. A more thorough study is needed to map these effects.

\section{Conclusions}
\label{conclusions}

We have studied initial conditions for inflation in scenarios where the inflaton potential has a plateau shape. Our motivation for this was two-fold: such models can be obtained in a large number of model classes and are, most importantly, among those most favored by Planck data. As a representative example, we considered a Higgs inflation model where the SM Higgs couples non-minimally to gravity, and also allowed the action to contain an $R^2$ term in the context of Palatini gravity. 

We showed that inflation with a large number of e-folds generically occurs in a large part of the model parameter space, without any need for fine-tuning of parameters even when the scale of inflation and the inflaton field value during inflation are much smaller than the Planck scale.  As shown in the paper, the phase-space trajectories end up in the regime where slow-roll inflation eventually takes place; either because the trajectories fall into the slow-roll regime directly, or because they pass $\chi=0$, slow down, and enter slow-roll on the other side of the potential. Only the trajectories with fine-tuned initial conditions end up oscillating around the origin without yielding a significant amount of slow-roll inflation. 

Our findings hold generically for all Higgs-like potentials where at least one of the parameters $\xi$, $\alpha_{\rm eff}$ is large. The above statements, however, depend on the choice of the coordinates and the corresponding phase-space probability measure, and should therefore be taken in a qualitative fashion only. Likewise, our conclusions apply only in the case where inflation occurs in a homogeneous and isotropic FRW universe. Future studies are needed to extend our study to scenarios where the initial conditions may not be homogeneous nor isotropic and to quantify the probability of inflation in such cases. 

Finally, we discussed the compatibility of the models studied in this paper with the CMB observations and the recently proposed trans-Planckian censorship conjecture, and showed that while the latter can easily be satisfied for large enough $\alpha_{\rm eff}$ (without fine-tuning of the initial conditions), the predicted value for the spectral index $n_s$ is, in this model, in tension with the Planck data. However, if the requirement of satisfying the conjecture is lifted, the models are in perfect agreement with data. This reinforces the motivation for future studies on this topic.


\section*{Acknowledgements}
We thank David I. Kaiser, Andrei Linde, and Ryan McManus for useful correspondence and discussions. TT is funded by the Simons foundation. ET acknowledges support from the Academy of Finland project 320123. This work was supported by the Estonian Research Council grants PRG803 and MOBTT5 and by the EU through the European Regional Development Fund CoE program TK133 ``The Dark Side of the Universe.''


\bibliography{Initial_conditions}


\end{document}